\documentstyle[graphicx]{mn}

\newif\ifAMStwofonts
\AMStwofontstrue

\def\mkn{{Mkn 3}}

\def\xmm{{\it XMM-Newton}}
\def\chandra{{\it Chandra}}

\def\et{{et al.\ }}

\def\ginga{{\it GINGA}}

\def\asca{{\it ASCA}}
\def\sax{{\it BeppoSAX}}

\newcommand{\ls}{\mathrel{\hbox{\rlap{\hbox{\lower4pt\hbox{$\sim$}}}\hbox{$<$}}}}
\newcommand{\gs}{\mathrel{\hbox{\rlap{\hbox{\lower4pt\hbox{$\sim$}}}\hbox{$>$}}}}

\def\arcs{{\hbox{$^{\prime\prime}$}}}

\def\Msun{\hbox{$\rm ~M_{\odot}$}}

\def\H0{{\rm ~km~s^{-1}~Mpc^{-1}}}

\def\et{{et al.}}

\def\deg{^\circ}

\title[X-ray spectrum of \mkn]
        {An \xmm\ observation of \mkn\ - a Seyfert galaxy just over the edge}
\author[K.A.Pounds \et]
        {K.A.Pounds,
	K.L.Page \\
Department of Physics and Astronomy, University of Leicester,
Leicester, LE1 7RH, UK\\}

\date{Accepted ; Submitted }
\pagerange{\pageref{firstpage}--\pageref{lastpage}}
\pubyear{2005}
\begin{document}
\maketitle
\label{firstpage}

\begin{abstract} 
A 100ks \xmm\ observation of the nearby Seyfert 2 galaxy \mkn\ offers a unique opportunity to explore the complexity of its
X-ray spectrum. We find the $\sim$3--8 keV continuum to be dominated by reflection from cold matter, with fluorescent K-shell
lines detected from Ni, Fe, Ca, Ar, S, Si and Mg. At higher energies an intrinsic power law continuum, with canonical
Seyfert 1 photon index, is seen through a near-Compton-thick cold absorber. A soft excess below $\sim$3 keV is found to
be dominated by line emission from an outflow of `warm' gas, photo-ionised and photo-excited by the intrinsically strong X-ray
continuum. Measured blue-shifts in the strong Fe K$\alpha$ and OVII and VIII emission lines are discussed in terms of the 
properties of the putative molecular torus and ionised outflow. 
\end{abstract}


\section{Introduction}
\mkn\ is a low redshift, $z=0.013509$ (Tifft and Cocke 1988), Seyfert 2 galaxy and one of the intrinsically brightest
AGN above $\sim$10 keV (Cappi \et\ 1999). However, absorption by a large column density of intervening cold matter (the
torus?) results in very little of the intrinsic X-ray continuum being directly visible. Instead, previous 
X-ray studies with \asca (Turner \et\ 1997a) and \sax (Cappi \et\ 1999) have suggested the observed X-flux over the 2-10 keV band 
is dominated by indirect
radiation `reflected' (ie re-directed by scattering in optically thick matter) into the line of sight. An extended soft X-ray
emission  component was resolved in an early \chandra\ observation (Sako \et\ 2000), which also showed the soft X-ray
spectrum to be dominated by blue-shifted line emission from highly ionised gas. \mkn\ is a particularly interesting object, being one of the 
small number of Seyfert 2
galaxies to exhibit broad optical emission lines in polarised light (Miller and Goodrich 1900, Tran 1995), while  the
observation of biconical [OIII] emission (Pogge and de Robertis 1993, Capetti \et\ 1995, 1999) also provides evidence
for an extended region of ionised gas. Soft X-ray spectra of \mkn\ have particular potential in studying an ionised
outflow in \mkn\ since it is one of the few bright Seyfert 2 galaxies not significantly `contaminated' by starburst X-ray
emission (Pogge and De Robertis 1993, Turner \et\ 1997b).

In this paper we present an analysis of archival data from an early \xmm\ observation of \mkn, which have not hitherto 
been published.

\section{Observation and data reduction}

\mkn\ was observed by \xmm\ on 2000 October 19-20 throughout orbit 158. We use X-ray spectra from the EPIC pn (Str\"{u}der \et 2001) and MOS (Turner
\et\ 2001) cameras, and the Reflection Grating Spectrometer/RGS (den Herder \et\ 2001). Both EPIC cameras were operated
in the full frame mode, together with the medium filter. The X-ray data were first screened with the SAS v6.1 software
and events corresponding to patterns 0-4 (single and double pixel events) were selected for the pn data and patterns
0-12 for the MOS data. EPIC source counts were taken within a circular region of 45\arcs\ radius about the centroid position of
\mkn, with the background being taken from a similar region, offset from but close to the source. The net exposures
available for spectral fitting were 76.5 ksec (pn), 136.5 ksec (combined MOS), 105.3 ksec (RGS1) and 102.0 ksec (RGS2).
Since no obvious variability was evident throughout the observation, spectral data were then integrated over the full
exposures and binned to a minimum of 20 counts per bin, to facilitate use of the $\chi^2$ minimalisation technique in
spectral fitting.  Spectral fitting was based on the Xspec package (Arnaud 1996)and  all spectral fits include
absorption due to the \mkn\ line-of-sight Galactic column of $N_{H}=8.7\times10^{20}\rm{cm}^{-2}$ (Stark \et\ 1992).
Errors are quoted at the 90\% confidence level ($\Delta \chi^{2}=2.7$ for one interesting parameter).

\begin{figure}                                                          
\centering                                                              
\includegraphics[width=6.3cm, angle=270]{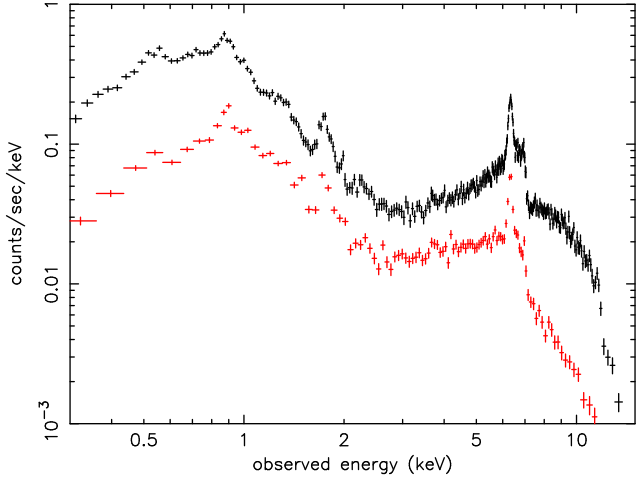}                     
\caption                                                                
{Background-subtracted data from the EPIC pn (black) and MOS (red) camera observations of \mkn\ showing the X-ray spectrum to
be detected over an unusually wide ($\sim$0.3--15 keV) spectral band}      
\end{figure}

\section{The hard (3-15 keV) EPIC spectrum} 

Figure 1 shows the background-subtracted EPIC camera data integrated over the full observation of \mkn. The spectrum is seen to be
highly structured and sufficiently hard to be detected to $\sim$15 keV (pn) and $\sim$11.5 keV (MOS). Given previous reports of a
physically separate soft spectral component below $\sim$3 keV we initially restricted modelling of the EPIC spectrum to the hard X-ray
(3--15 keV) band. A power law fit above 3 keV is poor but indicates the observed spectrum to be very hard, with a photon index
$\Gamma$$\sim$ -0.6. The more complex spectral model proposed by Turner \et (1997a) in analysing the \asca\ observation of \mkn, and
subsequently adopted by Cappi \et (1999) for their \sax\ analysis, was found to fit the data well. In that model the hard X-ray
continuum is a composite of a `canonical' Seyfert power law of photon index $\Gamma$$\sim$1.8, attenuated by a near-Compton-thick
absorber, and an unabsorbed reflection component. In addition the model contains an intense Fe K$\alpha$ line conceivably 
arising by
fluorescence in the same matter responsible for the continuum reflection.

In fitting this model to the EPIC data the principal power law and Fe K line parameters were left free and untied for the pn and MOS
fits. Cold reflection was modelled with PEXRAV in Xspec (Magdziarz and Zdziarski 1995), with a reflection factor R (=
$\Omega$/2$\pi$, where $\Omega$ is the solid angle subtended by the reflector at the continuum source) left free, but common to both
pn and MOS fits. The high energy cut-off of the incident power law was set at 200 keV (Cappi \et\ 1999), and the viewing angle (to
the reflector) at the default value of 60$\deg$. Element abundances were set to be solar. An excellent fit ($\chi^{2}$=1327 for 1336
degrees of freedom) was obtained for a photon index of $\Gamma$$\sim$1.82 (pn) and $\Gamma$$\sim$1.74 (MOS), attenuated by a column
of cold gas of N$_{H}$$\sim$$1.3\times 10^{24}$ cm$^{-2}$. The dominant reflection continuum was represented by R = 1.7$\pm$0.1. The
Fe K$\alpha$ emission line was well-fitted in the pn data by a gaussian at 6.417$\pm$0.004 keV (\mkn\ rest-frame), of width
$\sigma$=45$\pm$9 eV and flux=3.8$\pm$0.3$\times10^{-5}$ photons cm$^{-2}$ s$^{-1}$ . The corresponding Fe K$\alpha$ line parameters
from the MOS spectral fit were 6.417$\pm$0.004 keV, $\sigma$=38$\pm$11 eV and flux 4.7$\pm$0.4$\times10^{-5}$ photons cm$^{-2}$
s$^{-1}$.  The line equivalent width was 600$\pm$50 eV (pn) and  650$\pm$55 eV (MOS) compared with the ${\it total}$ local continuum.
This best-fit model for the 3--15 keV spectrum of \mkn\ is reproduced in figure 2. Compared only with the reflection continuum we
note the EW of the Fe K$\alpha$ line increases to 1050$\pm$80 eV(pn)and 1150$\pm$100 eV (MOS).

%plcabs will give Nh allowing for significant scattering

\begin{figure}                                                          
\centering                                                              
\includegraphics[width=6.3cm, angle=270]{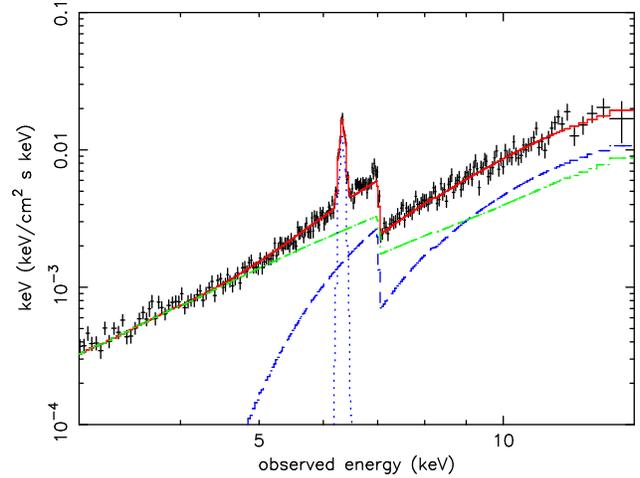}                     
\caption                                                                
{Unfolded model fit to the Mkn 3 EPIC spectrum $>$3 keV, as described in Section 3, showing the absorbed power law (dark blue) plus 
reflection (green) components to the hard continuum and a narrow Fe K$\alpha$ line (light blue).
The pn camera data only are shown for clarity}      
\end{figure}

Comparing the EPIC data with the above model in the neighbourhood of the Fe K$\alpha$ line (figure 3) reveals excess flux to higher 
energies, with a small peak near 7 keV.  Adding a second gaussian to the model finds a narrow line at 7.07$\pm$0.03 keV (AGN rest
frame), with a flux of 3.2$\pm$1.8$\times10^{-6}$ photons cm$^{-2}$ s$^{-1}$ and EW$\sim$110 eV (cf. the reflection continuum), an
addition which improves the fit to $\chi^{2}$=1301/1333 dof. Both the energy and relative strength are consistent with this being the Fe
K$\beta$ line, rather than FeXXVI Ly$\alpha$ proposed in the \asca\ analysis (Turner \et\ 1997a). The additional excess flux at
$\sim$6.5--6.8 keV was then modelled by a third narrow gaussian line at 6.69$\pm$0.05 keV (AGN rest frame), with a flux of 
$\sim$1.5$\times10^{-6}$ photons cm$^{-2}$ s$^{-1}$ and EW$\sim$40 eV (cf. the reflection continuum), further improving the fit to
$\chi^{2}$=1292/1330 dof. We note the energy of this emission line is close to that of He-like Fe. Finally, the small excess apparent
near 7.5 keV lies close to the energy of Ni K$\alpha$  emission. The addition of a fourth gaussian gave a further small improvement to
the fit ($\chi^{2}$=1285/1327 dof) with a line energy (rest frame) of 7.5$\pm$0.1 keV (compared with the laboratory energy of neutral Ni
K$\alpha$ of 7.47 keV), flux of  $\sim$2$\times10^{-6}$ photons cm$^{-2}$ s$^{-1}$ and EW$\sim$60 eV. Finally, we note figure 3 shows no
evident Compton shoulder on the low energy wing of the Fe K$\alpha$ line, which should occur by down-scattering when the re-processing
matter is optically thick to electron scattering. We return to this point in the next Section.

Based on the above broad band spectral fit we find an ${\it observed}$ 3-15 keV flux for \mkn\ of $1.5\times10^{-11}$ erg cm$^{-2}$
s$^{-1}$.  Correcting for the measured absorption the 3-15 keV flux increases to $4.8\times10^{-11}$ erg cm$^{-2}$ s$^{-1}$,
corresponding to an ${\it unabsorbed}$ luminosity of $1.9\times 10^{43}$ erg s$^{-1}$ ($ H_0 = 70 $ km\,s$^{-1}$\,Mpc$^{-1}$).
For later reference, extending that intrinsic power law spectrum to 0.3 keV would give a (0.3-15 keV) luminosity of 
$3.3\times 10^{43}$ erg s$^{-1}$.

\begin{figure}                                                          
\centering                                                              
\includegraphics[width=6.3cm, angle=270]{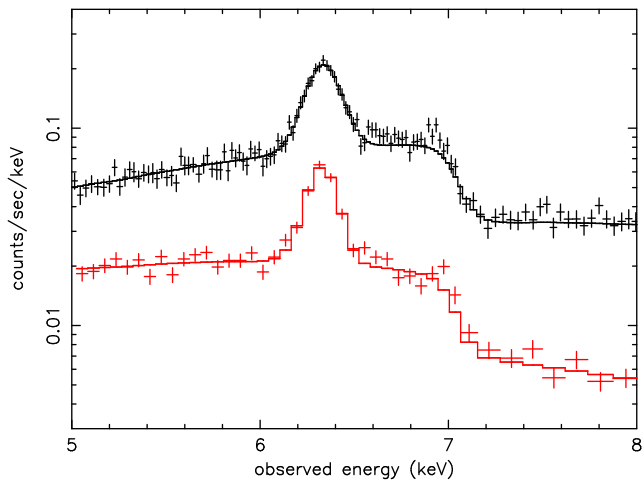}                     
\caption                                                                
{Comparison of the pn (black) and MOS (red) data with a power law + reflection + Fe K$\alpha$ line model, showing excess flux at $\sim$
6.5-6.8, 7 and 7.5 keV}      
\end{figure}

\subsection{A Compton shoulder to the Fe K$\alpha$ line}
The absence of an obvious Compton shoulder on the low energy wing of the strong Fe K$\alpha$ line is physically inconsistent with the
observed level of re-processing either by reflection or by transmission through a near-Compton-thick absorber (Matt 2002). The width of
the line is also a surprise if interpreted as due to velocity dispersion in matter on the far inner
face of the molecular torus. However, during the course of the present analysis we learned from MPE-MPG of a problem with pn counts
where the photon energy is deposited in two adjacent pixels, the so-called `doubles', whereby the registered photon energy can be
$\sim$20 eV greater than for the same photon fully absorbed in a single pixel (K Dennerl, private communication). To test the
importance of such effects on the present analysis we have repeated our fitting of the Fe K$\alpha$ line using only single pixel (pat0)
counts for the pn camera.

For the pn pat0 data we initially fitted a narrow gaussian line (ie with only the CCD resolution of $\sigma$=65 eV) to the Fe K$\alpha$
peak, allowing the energy and flux to be free. Figure 4 (top) shows how the narrow line then fits the high energy wing of the observed line
while leaving an excess on the low energy side. A second gaussian component was then added, at an energy $\sim$0.1 keV lower than the Fe
K$\alpha$ line, as expected for single Compton scattering in cold, optically thick matter (Matt 2002), and iterated with the Fe
K$\alpha$ line width and fluxes free.  While the double-gaussian fit (figure 4, bottom) was statistically no improvement on the original
single-gaussian fit to the Fe K$\alpha$ line, the relative strength of the Compton shoulder component to the primary Fe K$\alpha$ line
($\sim$20 percent) now gives a physically consistent overall picture of a reflection-dominated hard X-ray spectrum for \mkn. The above
procedure was then repeated for the MOS data, again finding an acceptable fit with a $\sim$ 25 percent Compton component to the Fe
K$\alpha$ line (figure 5).

\begin{figure}                                                          
\centering                                                              
\includegraphics[width=5.35cm, angle=270]{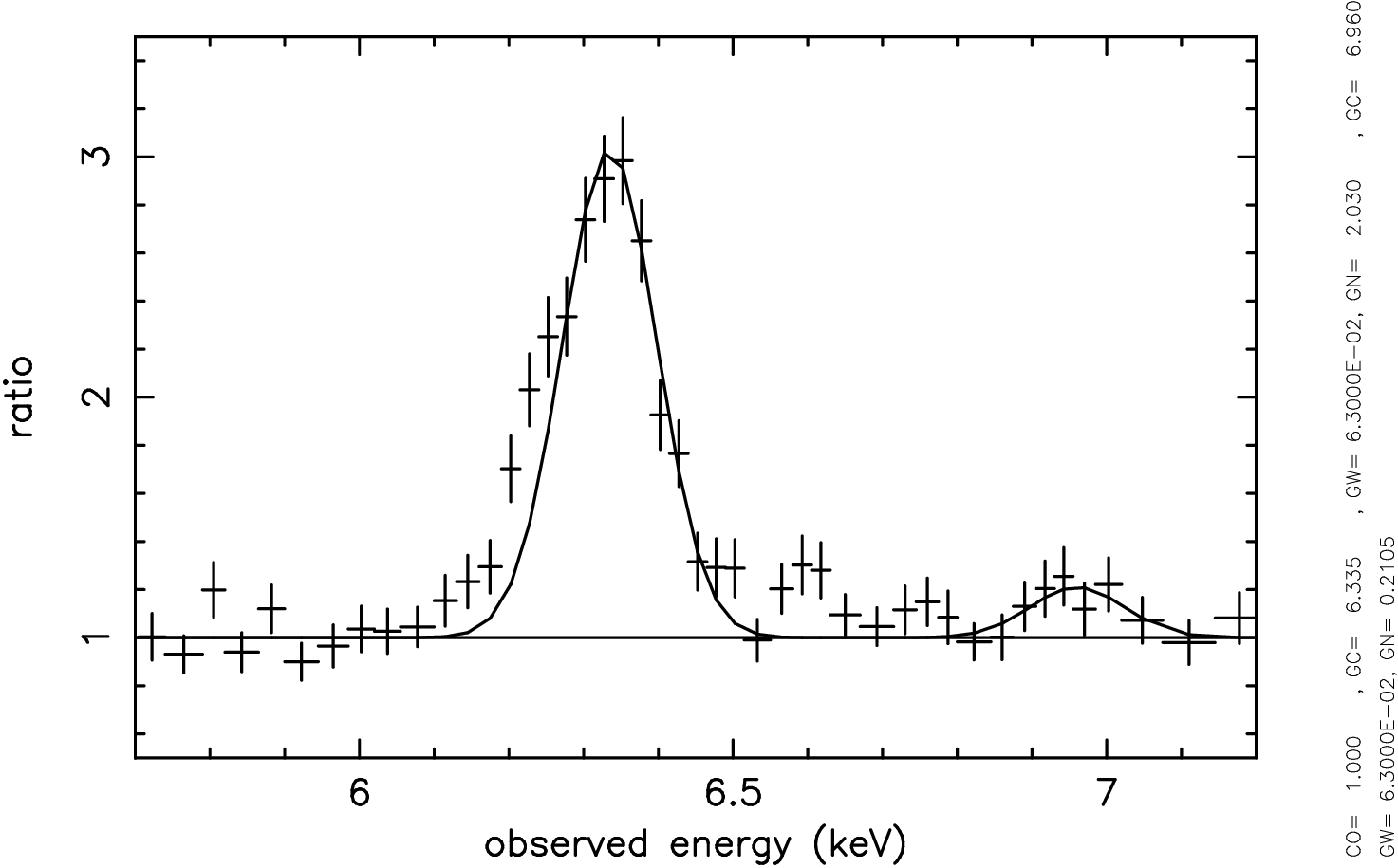}                     
\includegraphics[width=5.5cm, angle=270]{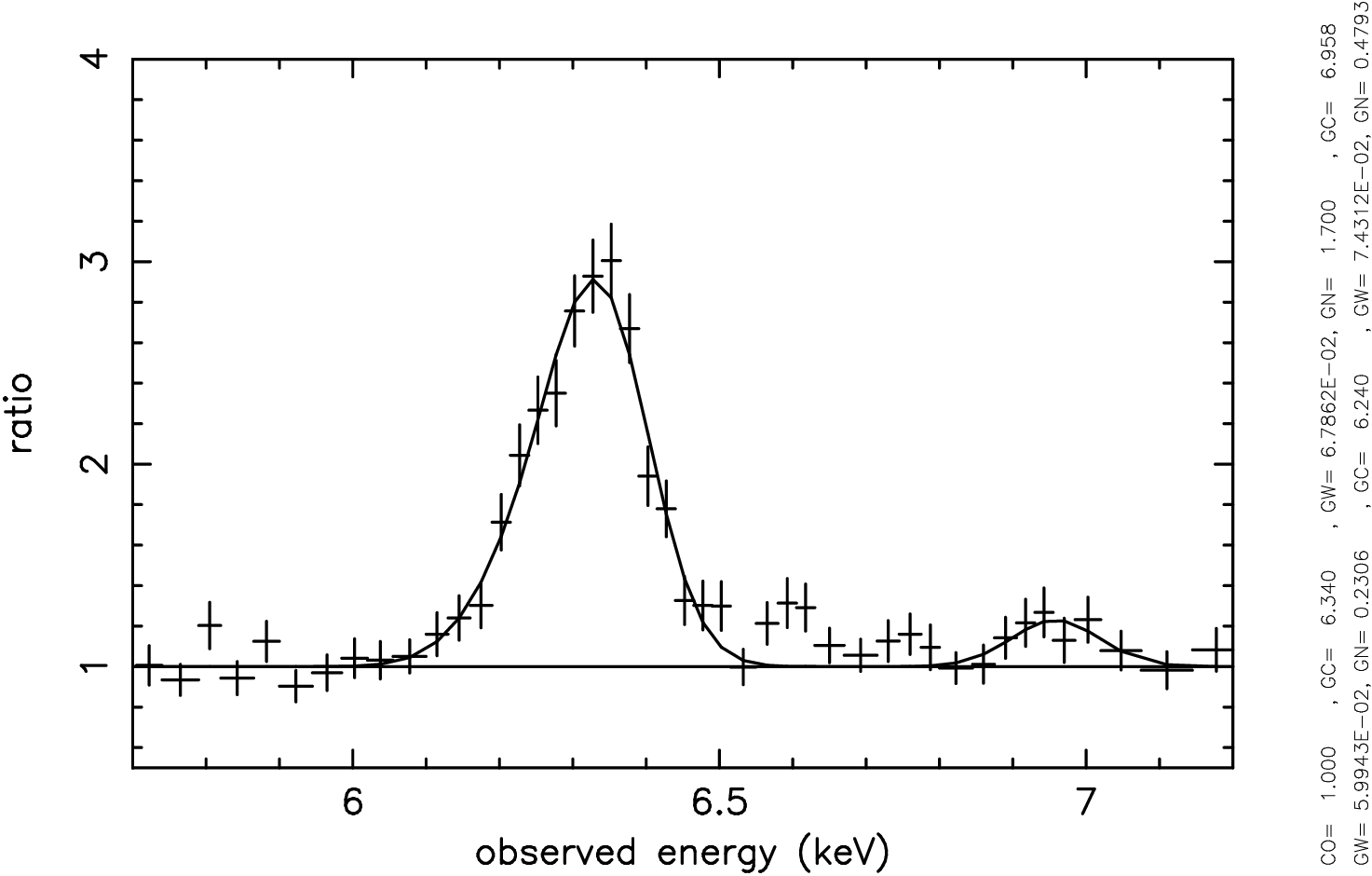}                     
\caption                                                                
{Gaussian lines added to the continuum model fit for the pn pat0 data: (top) narrow lines to model Fe K$\alpha$ and Fe K$\beta$
respectively; (bottom) third gaussian added to model the Compton scattering shoulder to the Fe K$\alpha$ line. Details in 
Section 3.1}      
\end{figure}                                                            

\begin{figure}                                                          
\centering                                                              
\includegraphics[width=5.5cm, angle=270]{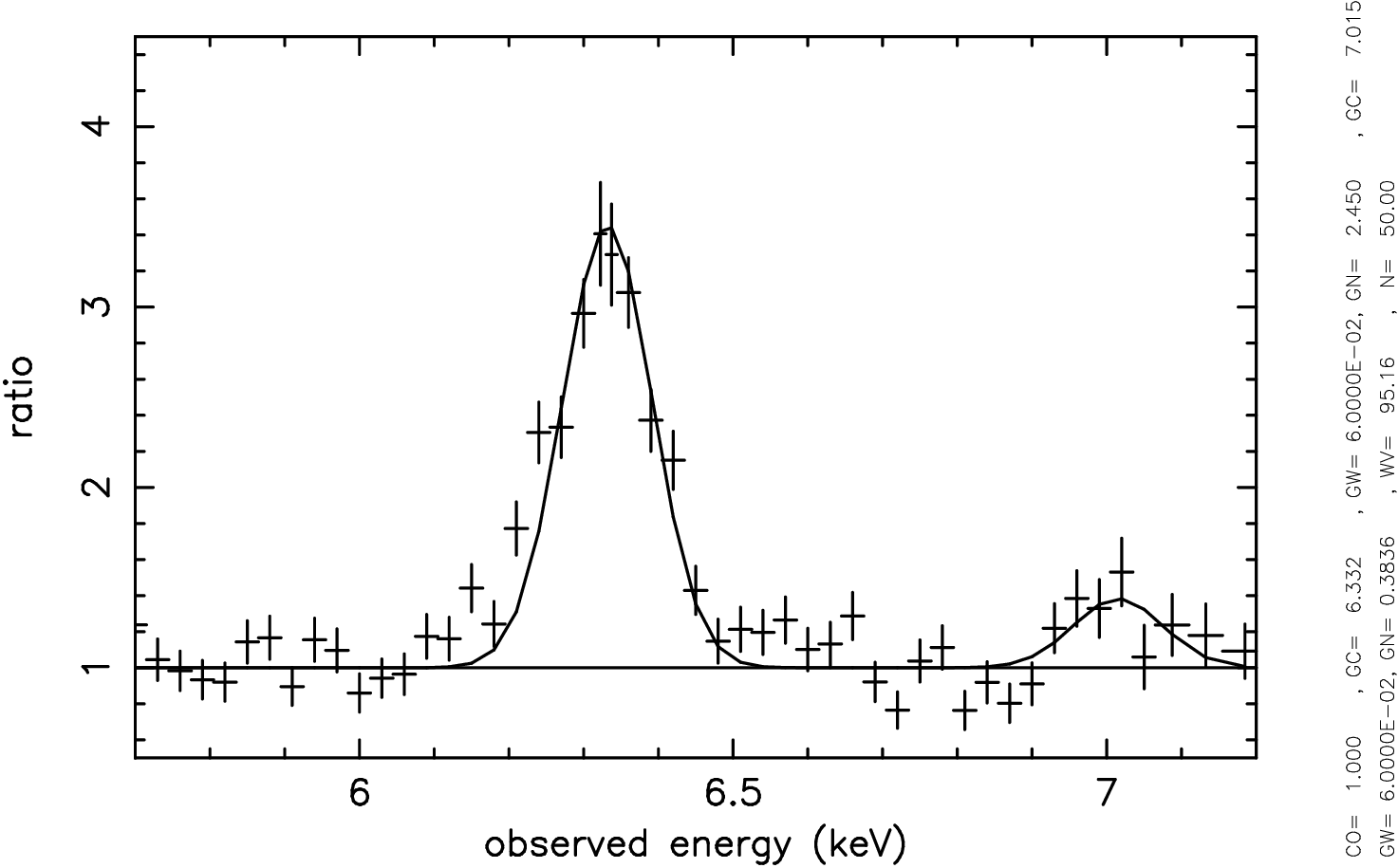}                     
\includegraphics[width=5.5cm, angle=270]{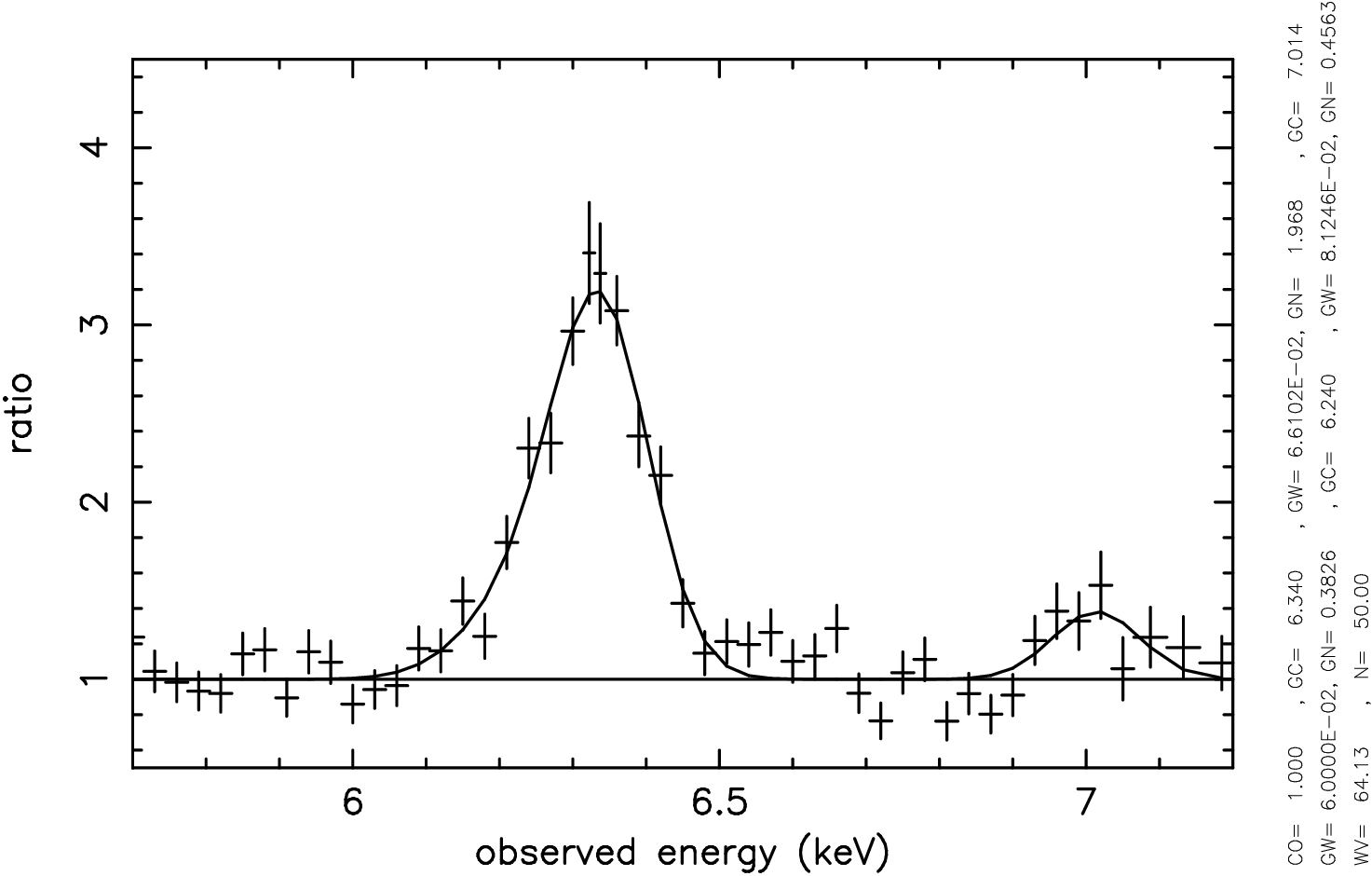}                     
\caption                                                                
{Gaussian lines added to the continuum model fit for the MOS data: (top) narrow lines to model Fe K$\alpha$ and Fe K$\beta$
respectively; (bottom) third gaussian added to model the Compton scattering shoulder to the Fe K$\alpha$ line. Details in 
Section 3.1}      
\end{figure}                                                            
 
\subsection{Fe K$\alpha$ line width and energy}
Allowing for the Compton shoulder has the important consequence that the primary Fe K$\alpha$ line is now only marginally resolved,
with a formal width of $\sigma$=30$\pm$15 eV in both pn and MOS fits. However, the primary line energy is ${\it increased}$, and though the
addition of the Compton component increases the errors, the primary line remains significantly blue-shifted, at 6.43$\pm$0.01
eV (\mkn\ rest-frame) for both pn pat0 and MOS data. Since fluorescent line emission will be produced in the cold reflector and the 
nuclear
absorbing column, it is reasonable to test for a common velocity. To do so we first fixed the Fe
K$\alpha$ line energy at 6.40 keV, the fit then worsening by $\Delta$$\chi^{2}$=43 for 1 additional dof. Allowing the redshift of the
line in Xspec to be free recovered the initial fit ($\chi^{2}$=1285/1327 dof), with the effective redshift falling (from z=0.0135) to
(1.07$\pm$0.05)$10^{-2}$. Tying the redshift of the cold absorber to that of the Fe K$\alpha$ line gave a further 
improvement to the 3--15 keV fit ($\Delta$$\chi^{2}$= 15) with a tied `redshift' of (1.05$\pm$0.04)$10^{-2}$. Tying the redshift of the
reflector to that of the line did not give a comparable improvement in the fit, though this was less well constrained. 
This test does suggest that a significant fraction of
the Fe K$\alpha$ line comes from the obscuring matter, which must then subtend a substantial solid angle to the primary source. That
geometry would, in turn, offer a possible velocity-dispersion explanation for the observed line width.

\subsection{Other spectral features in the EPIC data}

Extending the above hard X-ray (3-15 keV) model fit below 3 keV reveals a marked and highly structured `soft excess'.  This excess
remains (figure 6) after the
addition of a second (unabsorbed) power law component, with photon index fixed to the primary hard X-ray power law, intended to model
any continuum scattered into the line of sight by the ionised gas responsible for the polarised optical lines in \mkn. 
The inclusion of this unabsorbed power law component did, however, reduce
the cold reflection factor to R = 1.2$\pm$0.1. In the subsequent
EPIC analysis we retain this amended 3-15 keV spectral fit.

The soft excess illustrated in figure 6 is found to have an ${\it observed}$ flux (0.3-3 keV) of $4\times10^{-13}$~erg cm$^{-2}$ s$^{-1}$.
Allowing for the significant attenuation by the Galactic column of $N_{H}=8.7\times10^{20}$ cm$^{-2}$ increases this by $\sim$50 percent,
corresponding to an intrinsic soft X-ray luminosity over this energy band of $2.5\times 10^{41}$~erg s$^{-1}$ ($ H_0 = 70
$~km\,s$^{-1}$\,Mpc$^{-1}$). 

The reflection-dominated fit to the hard X-ray spectrum of \mkn\ suggests that fluorescent line emission from lighter, abundant metals
might be visible at lower energies, contributing to the spectral structure seen in figure 6. An initial examination of those pn
camera data shows that the strong peaks near 2.3 and 1.8 keV do lie close to the K$\alpha$ line energies of S and Si,  respectively. It is
important to note that these features are much stronger than any calibration uncertainties associated with the Si-K and Au-M edges in
XMM-EPIC. Similar structure is seen in the MOS spectrum and we proceed to study those data in more
detail, since in the soft X-ray band the MOS statistics are comparable to the pn camera while the energy resolution is better. [The 1$\sigma$ resolution of
the MOS at the  time of the \mkn\ observation was $\sim$34 eV at 1.5 keV (M.Kirsch, XMM-SOC-CAL-TN-0018 issue 2.3, 28 July 2004)].

\begin{figure}                                                          
\centering                                                              
\includegraphics[width=6.3cm, angle=270]{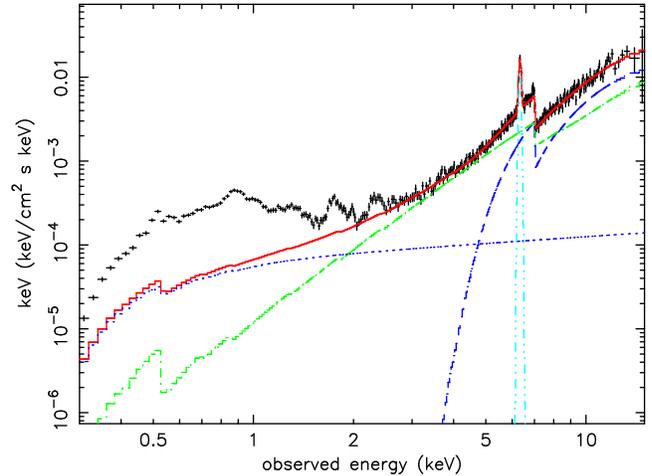}                     
\caption                                                                
{Low energy extension of the 3--15 keV spectral fit, revealing a soft excess. The addition of an unabsorbed continuum component
(blue dots)
with photon
index tied to the intrinsic power corresponds to a scattered fraction of the power law continuum. The pn data only
are shown for clarity.}      
\end{figure}

Closer examination of the low energy excess in the MOS spectrum confirms that the neutral K$\alpha$ lines of Si, S, Ca and
(probably) Ar are detected (figure 7). Table 1 compares the parameters of these lines, determined in Xspec, with their laboratory
energies and also with predicted equivalent widths for a cold solar abundance reflector (Matt \et\ 1997). The general agreement of
observed and predicted line strengths in Table 1 is good and provides further support for the reflection-dominated description
of the hard X-ray continuum of \mkn. 

In addition to revealing several fluorescent lines, the MOS spectrum of \mkn\ clearly resolves the principal resonance emission lines of
highly ionised ions of Si and S, as well as of Mg, Ne and O. The unusual richness of the EPIC spectral data at these low energies is a
consequence of the strong continuum source being hidden from direct view in \mkn. We examine this same spectral band with the higher
resolution RGS data in Section 4. 

\begin{figure}                                                          
\centering                                                              
\includegraphics[width=4.9 cm, angle=270]{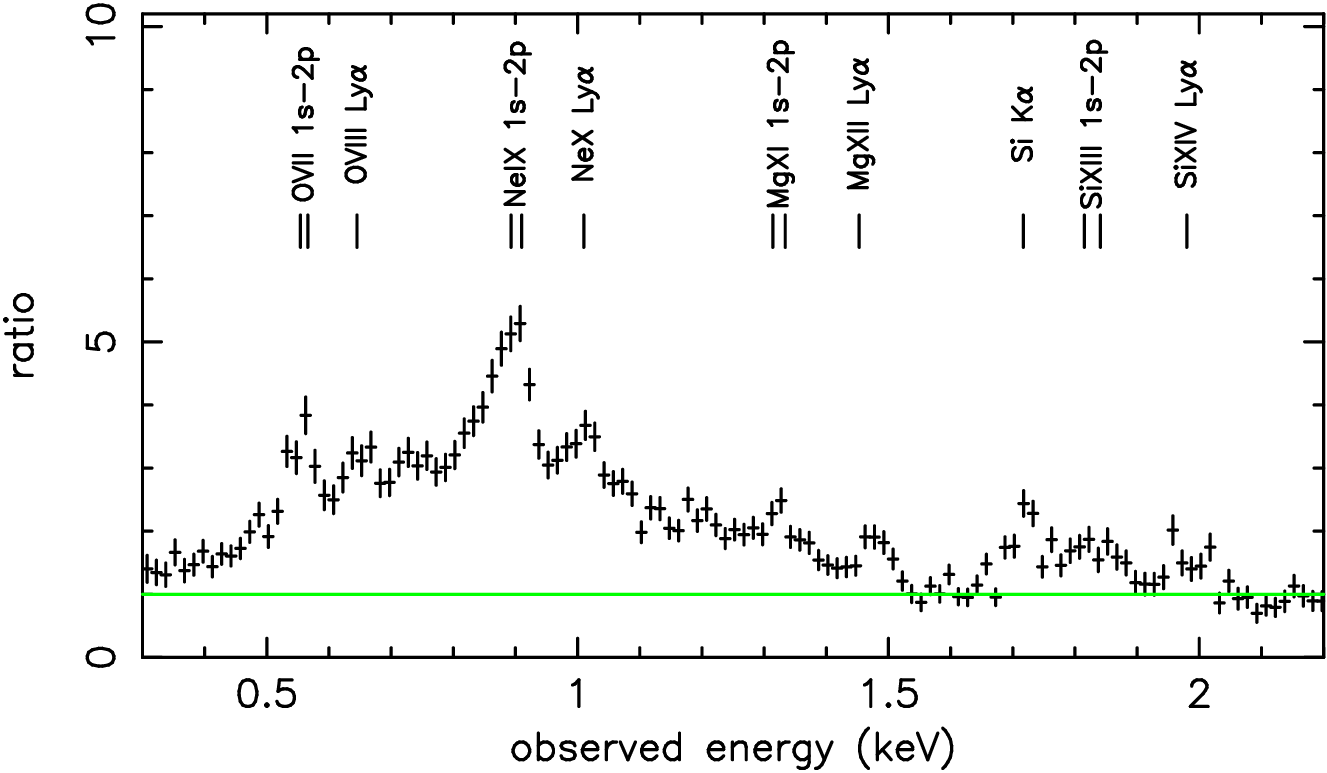}                     
\includegraphics[width=4.9 cm, angle=270]{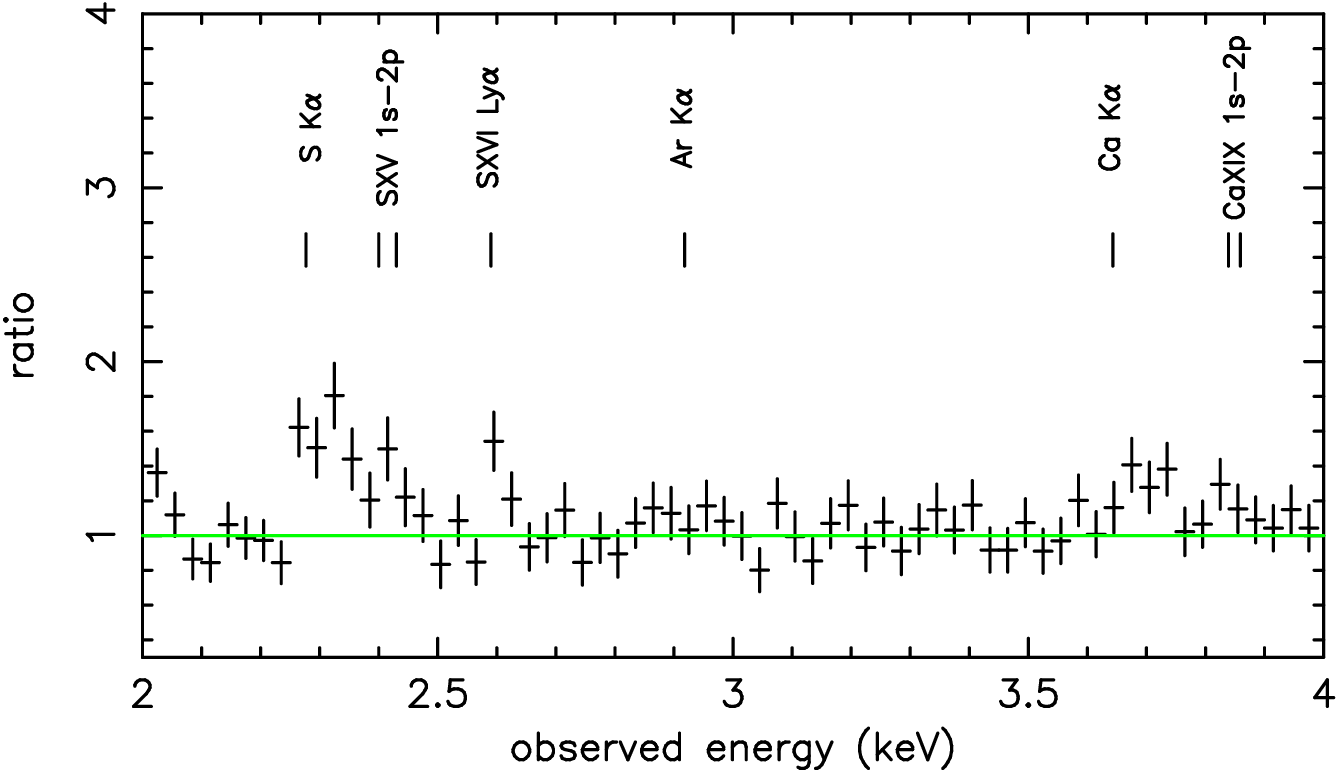}                     
\caption                                                                
{Identifying features in the EPIC (MOS camera) spectrum of \mkn\ with principal emission lines: (top) of 
highly ionised O, Ne, Mg, and Si, together with fluorescent Si K$\alpha$; (bottom) highly ionised S and Ca together 
with the K$\alpha$ lines of S, Ar and Ca}
\end{figure}

\begin{table*}
\centering
\caption{Candidate K$\alpha$ fluorescent emission lines in the EPIC spectrum of \mkn. The observed line energies (keV) 
are adjusted to the source  rest frame and compared with laboratory values. The observed equivalent widths (relative to
the reflection continuum) are compared with values estimated for solar abundances and a viewing angle of 60$\deg$ }

\begin{tabular}{@{}lccccc@{}}
\hline
Line &  E$_{source}$ & E$_{lab}$ & EW$_{obs}$(eV) & EW$_{est}$(eV) \\

\hline
Ni K$\alpha$ & 7.5$\pm$0.1 & 7.470 & 60 & 85 \\
Fe K$\alpha$ & 6.43$\pm$0.01 & 6.400 & 1100 & 1200 \\
Ca K$\alpha$ & 3.75$\pm$0.07 & 3.690 & 45 & 45  \\
Ar K$\alpha$ & 2.95$\pm$0.06 & 2.957 & 65 & 55  \\
S K$\alpha$ & 2.33$\pm$0.05 & 2.307 & 140 & 165  \\
Si K$\alpha$ & 1.75$\pm$0.05 & 1.740 & 200 & 230  \\

\hline
\end{tabular}
\end{table*}

\section{Spectral lines in the RGS data} 

The EPIC pn and MOS spectra of \mkn\ show a `soft excess' below $\sim$3 keV which contains both fluorescent emission from relatively cold
matter and line emission from highly ionised gas. We provisionally associate the latter with the extended soft X-ray
emission  resolved in the \chandra\ observation of \mkn\ (Sako \et\ 2000). To better quantify the soft X-ray spectrum of \mkn\ we
now examine the simultaneous \xmm\ grating data.

We find the soft X-ray spectrum of \mkn\ appears very similar to that of another nearby Seyfert 2 galaxy NGC 1068 (Kinkhabwala \et
2002), being dominated by line emission from highly ionised stages of the lighter metals. Unfortunately, the fluxes are an order of
magnitude less, mainly due to \mkn\ lying at a greater distance, but also being significantly attenuated at the longer wavelengths by
the relatively large Galactic column in the direction of \mkn. Nevertheless, visual examination of the RGS spectrum does allow the main
properties of an ionised outflow to be determined, adding to the picture reported by Sako \et\ (2000).

Figures 8--11 display the fluxed, background-subtracted RGS spectrum, combining both RGS-1 and RGS-2 data sets. The data are binned
at 40 m\AA, comparable to the intrinsic resolution of the RGS. It is immediately evident that the spectrum is dominated by line
emission and - except perhaps at the shortest wavelengths - the continuum level is low, consistent with the weak, unabsorbed
power law component in the EPIC spectral fit. 

\begin{figure} 
\centering 
\includegraphics[width=4.9 cm, angle=270]{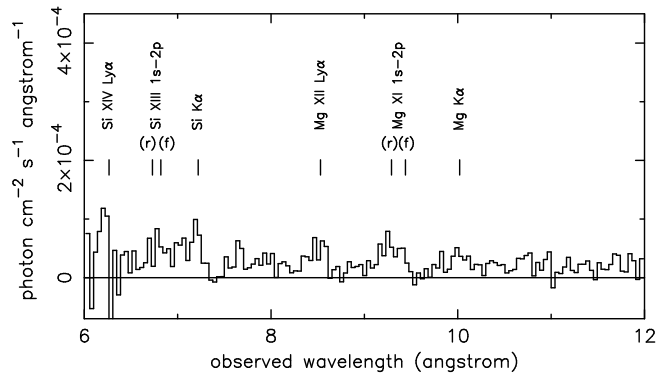} 
\caption {Fluxed RGS spectrum of \mkn\ over the waveband 6--12 \AA, binned at 
40 m\AA\ and showing 
emission lines identified with ionised and neutral Si and Mg}  
\end{figure}  

\begin{figure} 
\centering 
\includegraphics[width=4.9 cm, angle=270]{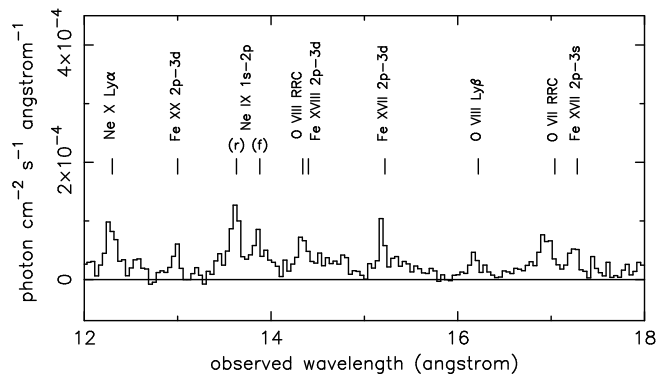} 
\caption {Fluxed RGS spectrum of \mkn\ over the waveband 12--18 \AA\ showing 
emission lines identified with ionised Ne, O and Fe, as well as RRC of OVII and OVIII}  
\end{figure}  

\begin{figure} 
\centering 
\includegraphics[width=4.9 cm, angle=270]{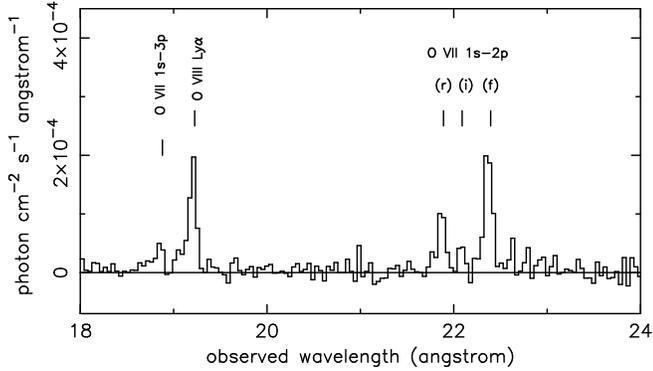} 
\caption {Fluxed RGS spectrum of \mkn\ over the waveband 18--24 \AA\ showing 
emission lines identified with OVIII Ly$\alpha$ and the OVII triplet}  
\end{figure}  

\begin{figure} 
\centering 
\includegraphics[width=4.9 cm, angle=270]{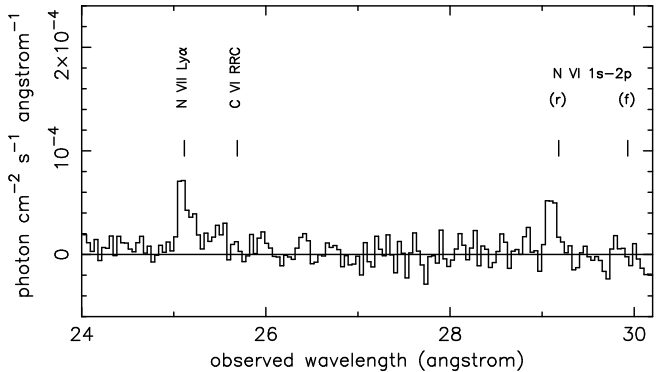} 
\caption {Fluxed RGS spectrum of \mkn\ over the waveband 24--30 \AA\ showing
emission lines identified with ionised N and C, together with a weak RRC of CVI}  
\end{figure}

\subsection{An emission line spectrum from a highly ionized gas.}
The shortest wavelength sections of the RGS spectrum overlap that resolved by the MOS camera and find good
qualitative agreement for the main emission lines of S, Si, and Mg. Furthermore, the MOS identifications of the strong 0.9
keV feature - and  the smaller peak near 1 keV - with the resonance lines of NeIX and NeX, respectively, are supported
by the higher resolution grating  spectra.

The RGS spectrum is characteristic of a recombining photoionised gas, the strongest emission lines being identified with
resonance transitions in He- and H-like ions of the most abundant elements for this waveband, viz. Si, Mg, Ne, O and N. The relative
weakness of Fe-L emission, together with the detection of narrow radiative recombination continuum (RRC) of OVII, OVIII and probably 
of CVI, indicates the gas is relatively cool (kT$\sim$6eV) and collisional ionisation/excitation unimportant. 

Marked on figures 8--11 are the laboratory wavelengths of the identified lines (adjusted for the redshift of \mkn), showing
the observed lines are generally shifted towards the blue. Table 2 lists the observed wavelengths and fluxes of the
principal lines.

To better quantify the blue-shift of the emission lines and also derive other parameters of the emitting plasma we fitted
gaussians to the fluxed spectra, now binned at 20m\AA , for the strong OVII triplet and the OVIII Ly$\alpha$ line (figures 12 and
13). The measured blue-shifts are: OVII-r (0.032$\pm$0.007\AA), OVII-f (0.037$\pm$0.005\AA) and OVIII
(0.029$\pm$0.005\AA), corresponding to line-of-sight velocities of 440$\pm$95, 500$\pm$70 and 450$\pm$75 km s$^{-1}$,
respectively. 

%should include uncertainties from RGS cal (see Kinkhabwala p.733)

\begin{figure} 
\centering 
\includegraphics[width=5.5 cm, angle=270]{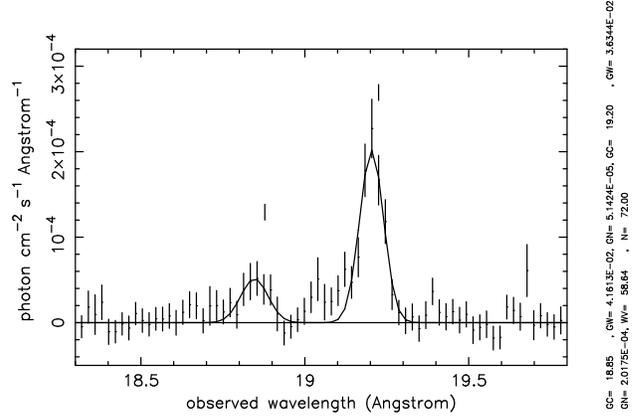} 
\caption  {Fluxed RGS spectrum of \mkn, binned at 20 m\AA, with gaussian fits to the 1s-3p emission of OVII and the Ly$\alpha$ line of 
OVIII}
\end{figure}  

\begin{figure} 
\centering 
\includegraphics[width=5.1 cm, angle=270]{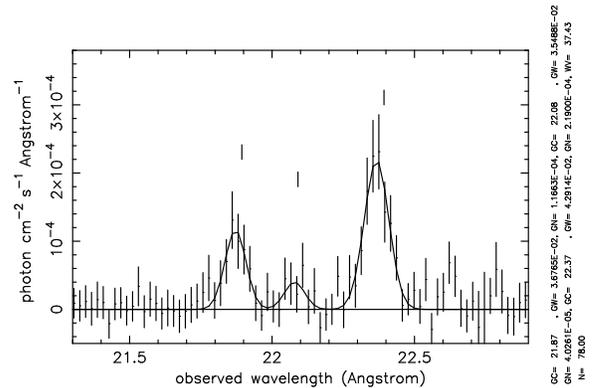} 
\caption  {Fluxed RGS spectrum of \mkn\ with gaussian fits to the resonance, intercombination and 
forbidden lines of OVII}  
\end{figure}

Assuming a common gaussian profile for the above lines yielded an observed width of 85$\pm$11 m\AA\ (FWHM). Allowing for
the RGS1 resolution of $60\pm$5 m\AA\ (FWHM), we find the OVII and VIII emission lines are broadened by $60\pm$12 m\AA\ (FWHM). 
Assuming velocity broadening that corresponds to a velocity dispersion of $900\pm$180 km s$^{-1}$.

In the OVII triplet (figure 12) the forbidden line is significantly stronger than the resonance line, with the intercombination lines
being relatively weak. While it is possible - perhaps probable - that the resonance line in OVII is affected by core absorption (see
Section 5.2) the strong forbidden line is consistent with emission from a photoionised gas in `the low density regime', viz
n$_{e}$$\leq$10$^{10}$ cm$^{-3}$  (Porquet and Dubau  2000). We discuss other information that can be deduced from the well-determined
OVII triplet in Section 5.2.

The suggestion of a `blue wing' to the stronger emission lines is interesting. This is most noticeable for
OVIII Ly$\alpha$ (figure 13), where there is no competing line blend, but can also be seen in the resonance lines of OVII and NeIX. 
If interpreted as a higher velocity component to the relevant emission line the corresponding projected outflow velocity would
be $\sim$3000 km s$^{-1}$. We return to this possibility in Section 6.

\begin{table*}
\centering
\caption{Principal emission lines identified in the RGS spectrum of \mkn. Wavelengths are in Angstroms
and adjusted to the source rest frame while line fluxes are in units of $10^{-6}$~photons cm$^{-2}$ s$^{-1}$.
The ionisation parameter corresponds to that giving a maximum abundance of the relevant ion calculated for an optically thin gas
irradiated by a $\Gamma$=1.8 power law using XSTAR}

\begin{tabular}{@{}lcccccc@{}}
\hline
Line &  $\lambda$$_{source}$ & $\lambda$$_{lab}$ &  flux  &  log$\xi$ \\

\hline
SiXIV Ly$\alpha$ & 6.16$\pm$0.04 & 6.18 & 8$\pm$2 & 2.6 \\
SiXIII 1s-2p (r) & 6.62$\pm$0.07 & 6.65 & 4$\pm$2  & 2.1 \\
SiXIII 1s-2p (f) & 6.71$\pm$0.05 & 6.74 & 6$\pm$2  & 2.1 \\
Si K$\alpha$ & 7.11$\pm$0.03 & 7.13 & 10$\pm$2  & - \\
MgXII Ly$\alpha$ & 8.39$\pm$0.05 & 8.42 & 8$\pm$3 & 2.3 \\
MgXI 1s-2p (r) & 9.14$\pm$0.03 & 9.17 & 7$\pm$2  & 2.0 \\
MgXI 1s-2p (f) & 9.27$\pm$0.05 & 9.31 & 5$\pm$2  & 2.0 \\
Mg K$\alpha$ & 9.87$\pm$0.04 & 9.89 & 5$\pm$2  & - \\
NeX Ly$\alpha$ & 12.11$\pm$0.03 & 12.13 & 9$\pm$2 & 2.0 \\
FeXX 2p-3d & 12.83$\pm$0.02 & 12.84 & 4$\pm$2  & 2.4 \\
NeIX 1s-2p (r) & 13.43$\pm$0.03 & 13.45 & 8$\pm$2 &  1.5 \\
NeIX 1s-2p (f) & 13.68 $\pm$0.05 & 13.70 & 6$\pm$2 &  1.5 \\
FeXVII 2p-3d & 14.98$\pm$0.04 & 15.01 & 7$\pm$2 &  1.8 \\
OVIII Ly$\beta$ & 15.97$\pm$0.03 & 16.01 & 5$\pm$1 &  1.6 \\
FeXVII 2p-3s & 17.03 $\pm$0.03 & 17.06 & 7$\pm$2  &  1.8 \\
OVII 1s-3p & 18.60$\pm$0.03 & 18.63 & 5$\pm$1   &  1.0 \\
OVIII Ly$\alpha$ & 18.940 $\pm$0.005 & 18.969 & 22$\pm$3 & 1.6  \\
OVII 1s-2p (r) & 21.570 $\pm$0.007 & 21.602 & 9$\pm$2  &  1.0 \\
OVII 1s-2p (i) & 21.78 $\pm$0.04 & 21.80 & 4$\pm$1  &  1.0 \\
OVII 1s-2p (f) & 22.058$\pm$0.005 & 22.095 & 22$\pm$4   &  1.0 \\
NVII Ly$\alpha$ & 24.77$\pm$0.03 & 24.78 & 9$\pm$2  &  1.3 \\
NVI 1s-2p (r) & 28.71$\pm$0.05 & 28.80 & 8$\pm$3   &  0.7 \\

\hline
\end{tabular}
\end{table*}

\section{Discussion}

\subsection{A reflection-dominated hard X-ray spectrum}

Our analysis of the EPIC data confirms and refines the model proposed by Turner \et (1997) and later adopted by Cappi \et (1999) in
their respective analyses of \asca\ and \sax\ observations of \mkn. The improved quality of the EPIC spectra are significant in several
respects. First, observing the hard X-ray spectrum to $\sim$15 keV allows the absorbed power law and reflection components to be
separated and quantified, for the first time within a single data set. This confirms \mkn, one of the brightest AGN in the hard X-ray
sky, to have an intrinsic power law continuum typical of a Seyfert 1 nucleus, but attenuated by a near-Compton-thick absorber of cold
matter, resulting in the observed X-radiation at $\sim$3--8 keV being dominated by cold reflection. Second, the fluorescence line
emission of all abundant elements from Ni to Si (and possibly Mg) are detected, consistent with the reflection-dominated model. Third,
the Fe K$\alpha$ line is sufficiently well determined to indicate a small but significant `blue-shift'. We find no evidence for an
absorption edge at $\sim$8.4 keV reported in an independent analysis of the \asca\ data by Griffiths \et (1998). Conversely, the edge
energies for both the absorber and reflector lie close to 7.1 keV, appropriate to near-neutral Fe. 

Removing the pn double-pixel counts, and including a Compton shoulder, reduces the apparent width of the Fe K$\alpha$ line in the pn
data to a `marginally resolved' $\sigma$=32$\pm$15 eV. Allowing for the Compton shoulder in the MOS fit makes the line width similarly
marginal. However, inclusion of the Compton scattered component ${\it increases}$ the energy of the primary line, which now lies in the
range 6.43$\pm$0.01 keV (at the redshift of \mkn) in both pn and MOS data, compared with a laboratory energy (weighted mean  of
K$\alpha$$_{1}$ and K$\alpha$$_{2}$) of 6.40 keV. With an absolute error of $\sim$5 eV for the MOS and $\sim$10 eV for the pn camera
(XMM-SOC-CAL-TN-0018), we conclude the line is `blue-shifted' by 30$\pm$10 eV.  If that shift were due to the ionisation state of the
reflecting matter it would correspond to Fe XV - XVIII (House 1969); however that would show the Fe K edge energy $\ga$7.5 keV, in
conflict with the value observed. The most obvious alternative is a velocity shift with a projected (line of sight) velocity of
3200$\pm$1100 km s$^{-1}$.  We note that neither the line energy nor (marginally resolved) line width of Fe K$\alpha$ in \mkn\ sit
comfortably with cold reflection from the unobscured far-side inner wall of the putative molecular torus. Interestingly, Iwasawa et al
(1994) also found a blue-shifted Fe K$\alpha$ line in the \asca\ spectrum, and - furthermore - claimed the line flux to vary over a 3.5
year interval. Griffiths \et (1998) also concluded the Fe K$\alpha$ line flux fell by a factor $\sim$1.8 between \ginga\ and \asca\
observations in 1989 and 1993. On the other hand Sako et al (2000) reported a K$\alpha$ line at 6.391$\pm$0.004 keV (before allowing for any
Compton shoulder) from an early \chandra\ HETG observation, with a width of $\sim$36$\pm$4 eV.

Taken overall, the form of the `cold' reprocessor indicated by the EPIC spectra is unclear. The strong reflection continuum and $\sim$1
keV EW Fe K$\alpha$ line match well with an origin in the exposed far inner wall of a torus, the near side of
which forms the absorbing barrier to the primary power law source. However, the observed Fe K$\alpha$ line energy  - and similar
blue-shift of the Fe K edge in the absorber - suggests a significant
part of the fluorescent emission
may arise 
in cold, dense `clouds'
in the same `near-side' outflow that also obscures the continuum source from direct view in \mkn. 

\subsection{Extended photoionised gas in \mkn.}

The RGS spectrum of \mkn\ is dominated by narrow emission lines and RRC of highly ionised gas. The agreement with the \chandra\
observation reported by Sako \et\ (2000) is good, consistent with the conclusion of those authors that both photoionisation/recombination
and photo-excitation/radiative emission are important in the soft X-ray spectrum of \mkn.  While not affecting that basic conclusion the 
independent determination of line energies and fluxes with the RGS is useful, given the low count rates in both observations.
The characteristics of the RGS and \chandra\ HETGS are also complementary, with the former having higher sensitivity for the important
OVII and VIII lines (with $\sim$100-200 counts per line) while the HETGS extends the high resolution spectrum to much higher energies. 
The RGS data longward of $\sim$ 12 \AA\ shows the underlying continuum to be
negligible, consistent with the broad-band EPIC fit (eg figure 3) which includes a (maximum) power law component of $\sim$1 percent of
the intrinsic power law. If electron scattering from the ionised outflow produces the polarised broad emission lines 
seen in the optical spectrum of \mkn\
(Miller and Goodrich 1990, Tran 1995), our upper limit to the unabsorbed power law component constrains the product C$\tau$ to be
$\leq$0.01, where C is the covering factor of the ionised outflow and $\tau$ its optical depth in the line of sight.

Further information on the ionised gas can be obtained from the emission line spectrum. The observed intensity of the
forbidden (22.1 \AA) line of OVII is a useful measure, since it is unaffected by resonance absorption. Assuming a solar 
abundance of oxygen, with 25 percent in OVII, 50 percent of recombinations from OVIII direct to the ground state, and a
recombination rate at kT $\sim$6 eV (from the corresponding RRC width) of $10^{-11}$~cm$^{3}$~s$^{-1}$ (Verner and
Ferland 1996), we estimate an emission measure of order $1.5\times$$10^{64}$ cm$^{-3}$. The ion species listed in Table
3 suggest the co-existence of gas over a wide range of ionisation, yielding a total emission measure a factor several higher. The extent
of the emission region is not directly constrained by the line intensities or ionisation parameter, but given the
Seyfert 2 nature of \mkn\ a reasonable minimum radius is r$\sim$1 pc ($3\times10^{18}$ cm). With the emission volume approximated 
by a cylinder of radius r and length 10r, a uniform density gas would then have n$_{e}$$\sim$$4\times$$10^{3}$ cm$^{-3}$.

The ionisation parameter $\xi$ (= L/nr$^{2}$) provides a check on those simple estimates, where L  is the (unabsorbed) X-ray continuum
flux irradiating the outflow. The EPIC spectral fit found L($\ga$0.3 keV)$\sim$$2.6\times 10^{43}$ erg s$^{-1}$, giving an ionisation
parameter $\xi$$\sim$30 erg cm s$^{-1}$ at r=5 pc, a value at which OVII  and OVIII would both be prominent. In fact, the detection of
He-like NVI and H-like SiXIV  in the RGS spectrum shows the emitting gas must cover a wide range of ionisation parameter, those ions
having peak abundances at $\xi$ of $\sim$5 erg cm s$^{-1}$ and $\sim$400 erg cm s$^{-1}$, respectively. Assuming the ionised Si and Ca emission lines detected in the MOS spectrum
arise in the same outflow requires still more highly ionised gas, while the excess flux in the EPIC data at $\sim$6.6-6.8keV, if
correctly attributed to emission from FeXXV, extends that range to $\xi$$\sim$1000 erg cm s$^{-1}$. Such a broad ionisation structure, where the
distance to the ionising source is constrained to r$\ga$r$_{BLR}$, suggests an inhomogeneous gas with a correspondingly broad range of
density. While the present spectra are clearly inadequate to allow such a complex ionised outflow  to be mapped it is possible to
estimate several additional parameters. 

Line ratios in the He-like triplet are an establised diagnostic of ionised plasmas. In the RGS spectrum of \mkn\ the OVII triplet  is
well defined, with the forbidden (f), resonance (r) and intercombination (i) lines all detected (figure 12). The observed line strengths
are in the ratio 22:9:4 yielding values of the diagnostic parameters R(=f/r+i)$\sim$5 and G(=f+i/r)$\sim$3, rather high and low,
respectively, compared with the calculated ratios for a pure, low density photoionised plasma (Porquet and Dubau 2000). More  recent
calculations, which take account of the column density and ionisation parameter in the radiating gas (Godet \et\ 2004), yield G=3 for
OVII for a column density of order N$_{H}$$\sim$$10^{21-22}$ cm$^{-2}$ and $\xi$ = 30-100 erg cm s$^{-1}$.  Encouragingly, that column is of the same
order as the product n$_{e}$r in the above simple fit to the calculated emission measure, and also to the estimated scattering $\tau$
for a covering factor of 0.1. In particular the resonance line of OVII (21.6 \AA) will be optically thick in the core, limiting the
effects of photo-excitation in comparison with higher order lines. We note also that the observed f:r line ratios are lower in the Ne and Mg
triplets, suggesting the opacity effects there are less. 

The mass of the (observed) extended gas envelope is $\sim$900\Msun, assuming a uniform density.  With a projected outflow velocity,
observed at
$\sim$45 degrees to the line of sight, of $\sim$700 km s$^{-1}$,  the mass outflow rate is then $\sim$0.04 \Msun yr$^{-1}$, with an
associated mechanical energy of $2\times10^{40}$ erg s$^{-1}$. These estimates would be reduced if the emitting gas is
indeed clumpy.

From the absorption-corrected 2--10 keV luminosity of $1.5\times 10^{43}$~erg s$^{-1}$ we estimate a bolometric luminosity for \mkn\ of
$4\times 10^{44}$~erg s$^{-1}$. At an accretion efficiency of 0.1 the accretion mass rate is then $\sim$0.06 \Msun  yr$^{-1}$. Thus, as
may generally be the case in Seyfert 1 galaxies, we find the ionised outflow in \mkn\ carries a significant mass loss from the AGN.
However, at least by a radial distance r$\ga$r$_{BLR}$, the kinetic energy in the outflow is relatively small, even compared
to the soft X-ray luminosity. 

\section{Conclusions}

\xmm\ EPIC observations of high statistical quality confirm previous findings that the extremely hard (2--10 keV) 
power law index in the Seyfert 2 galaxy \mkn\ is due to strong continuum reflection from `cold' matter. The intrinsic
power law, which is seen through a near-Compton-thick absorber, emerges above $\sim$8 keV and has a photon index typical of 
Seyfert 1 galaxies. The dominant reflection component in \mkn\ appears to be unaffected by the large absorbing column,
allowing fluorescent line emission to be detected from Ni K$\alpha$ ($\sim$7.5 keV) down to Mg K$\alpha$ ($\sim$1.25keV).

The combination of a large column density obscuring the continuum source with the visibility of a large area of cold reflector 
suggests that \mkn\ is being viewed at an inclination which cuts the edge of the obscuring screen. This would be consistent
with the geometry of `Seyfert galaxies on the edge', identified in a recent extensive review of the optical polarisation properties 
of Seyfert galaxies (Smith \et\ 2002, 2004). Comparison with the \xmm\ spectrum of NGC4051 (eg Schurch \et\
2003) suggests that well-studied Seyfert 1 galaxy may be inherently similar but is being viewed from the other (lower obscuration) side
of a (blurred) edge.

The strong Fe K$\alpha$ line appears to be `blue-shifted' in both EPIC pn and MOS spectra, indicating an origin in low ionisation matter
with a projected outflow velocity of 3100$\pm$1100 km s$^{-1}$. Although less well resolved after the inclusion of a Compton shoulder,
the Fe K$\alpha$ line has a formal width of 3200$\pm$1600 km s$^{-1}$ (FWHM). The projected velocity and indicated velocity dispersion
are both inconsistent with an origin at the far inner wall of a torus of r$\ga$1 pc. It seems more likely that a large fraction of the
neutral K$\alpha$ emission arises in dense matter circulating or outflowing at a radial distance more typical of the BLR clouds. For
comparison, we note that Tran
(1995) found the FWHM of H$\beta$ in polarised light to be 6000 km s$^{-1}$ in \mkn. 

Below $\sim$3 keV a `soft excess' emerges above the hard, reflection-dominated continuum. Spectral structure is resolved
in both EPIC and RGS data, with the MOS spectrum showing remarkable detail. In this soft X-ray band the spectrum is
found to be dominated by resonance lines of He- and H-like Si, Mg, Ne, O and N, with an observed blue-shift of 470$\pm$70
km s$^{-1}$, indicating an origin in a highly ionised outflow extending above the Seyfert 2 absorbing screen. 

Relative line fluxes and the detection of narrow (low temperature) radiative recombination continua of OVII, OVIII and (probably) CVI
are all consistent with the gas being photo-ionised/photo-excited by the intrinsic power law continuum, as found by Sako \et\ (2000). In
all these respects the soft X-ray emitting gas in \mkn\ is very similar to that seen in another nearby, and still brighter, Seyfert 2
galaxy NGC1068 (Kinkhabwala \et\ 2002). As those authors pointed out, the ionised outflow seen in emission in NGC1068 - and in \mkn\ -
is  consistent with that  typically seen in absorption against the power law continuum in Seyfert 1 galaxies. The soft X-ray
luminosity in \mkn\ is of order 1\% of the intrinsic (absorption corrected) 0.3-10 keV luminosity.

It is interesting to compare that figure with the `non-varying' component of the soft X-ray excess found in the Seyfert 1 galaxies 
(NGC 4051, Pounds \et\ 2004a) and 1H0419-577 (Pounds
\et\ 2004b), where soft X-ray luminosities of 5\% and 9\% of the dominant power law component were indicated. It was suggested in those
papers that a large part of the non-varying soft flux might be explained by the same outflow usually seen in absorption. 
For that explanation to hold up it would appear that a bright inner region, shielded from direct view in a
Seyfert 2 such as \mkn, must have a sufficiently high velocity dispersion to remain undetected in high resolution absorption spectra.
Interestingly, Gierlinski and Done outlined an extreme case of such a scenario in proposing an absorption-based alternative to the strong
soft excess in the luminous Seyfert 1 galaxy PG1211+577 (Gierlinski and Done, 2004).
The blue wings seen on some of the stronger emission lines in the RGS spectrum may be an indication of such a trend to higher velocity 
gas
extending into the unobscured outflow in \mkn. However, significant emission could also come from a second ionised emission component
out of the line-of-sight to the continuum source (and hence not seen in absorption in Seyfert 1s) within the cone of the obscuring
torus. Such a component might be identified with the `equatorial scatterer' required by optical polarisation studies 
of Seyfert galaxies (Smith \et\ 2002, 2004).

\section*{ Acknowledgements } The results reported here are based on observations obtained with \xmm, an ESA science
mission with instruments and contributions directly funded by ESA Member States and the USA (NASA). The authors wish to
thank Leicester colleagues for valuable input, the SOC and SSC teams for organising the \xmm\ observations and initial data
reduction. KAP acknowledges the support of a Leverhulme Trust Emeritus Fellowship and KPA of a PPARC research grant.

\end{document}